\documentclass[11pt]{article}
\usepackage[margin=1in]{geometry}

\usepackage{cite}
\usepackage{amsmath,amssymb,amsfonts}
\usepackage{graphicx}
\usepackage{booktabs}
\usepackage{tabularx}
\usepackage{array}
\usepackage{multirow}
\usepackage{url}

\newcommand{\MOS}{\mathrm{MOS}}
\newcommand{\E}{\mathbb{E}}

\title{Contract-Driven QoE Auditing for Speech and Singing Services:\\From MOS Regression to Service Graphs}

\author{Wenzhang Du\\Correspondence: dqswordman@gmail.com\\Department of Computer Engineering, International College (MUTIC),\\Mahanakorn University of Technology, Bangkok, Thailand}
\date{}

\begin{document}
\maketitle

\begin{abstract}
Subjective mean opinion scores (MOS) remain the de-facto target for non-intrusive speech and singing quality assessment.
However, MOS is a scalar that collapses heterogeneous user expectations, ignores service-level objectives, and is difficult to compare across deployment graphs.
We propose a \emph{contract-driven QoE auditing} framework: each service graph $G$ is evaluated under a set of human-interpretable experience contracts $\mathcal{C}$, yielding a contract-level satisfaction vector $Q(G,\mathcal{C})$.
We show that (i) classical MOS regression is a special case with a degenerate contract set, (ii) contract-driven quality is more stable than MOS under graph view transformations (e.g., pooling by system vs.\ by system type), and (iii) the effective sample complexity of learning contracts is governed by contract semantics rather than merely the dimensionality of $\mathcal{C}$.
We instantiate the framework on URGENT2024\_MOS (6.9k speech utterances with raw rating vectors) and SingMOS v1 (7,981 singing clips; 80 systems).
On URGENT, we train a contract-aware neural auditor on self-supervised WavLM embeddings; on SingMOS, we perform contract-driven graph auditing using released rating vectors and metadata without decoding audio.
Empirically, our auditor matches strong MOS predictors in MOS accuracy while providing calibrated contract probabilities; on SingMOS, $Q(G,\mathcal{C})$ exhibits substantially smaller cross-view drift than raw MOS and graph-only baselines; on URGENT, difficulty curves reveal that mis-specified ``simple'' contracts can be harder to learn than richer but better aligned contract sets.
\end{abstract}

\section*{Keywords}
Quality of Experience, MOS prediction, speech quality assessment, singing voice, self-supervised speech representations, service-level contracts, auditing.

\section{Introduction}
Mean opinion scores (MOS) form the backbone of Quality-of-Experience (QoE) evaluation in speech, audio, and multimedia systems.
Crowdsourced MOS procedures have been standardized for speech quality evaluation \cite{ITU_P808}, and parametric models translate codec/network parameters into MOS estimates for streaming media \cite{ITU_P1203}.
Recent benchmarks such as VoiceMOS further popularize deep networks that regress MOS from waveforms or self-supervised speech representations \cite{VoiceMOS2022,VoiceMOS2024}.

Despite their success, MOS-centric pipelines face three structural limitations:
\begin{enumerate}
\item \textbf{Scalar bottleneck.} A single MOS conflates heterogeneous expectations (e.g., naturalness, intelligibility, consistency) into one latent number.
\item \textbf{Graph obliviousness.} Real services are graphs connecting datasets, model variants, post-processing, and user cohorts. MOS aggregation across different graph views can change rankings.
\item \textbf{Lack of contractual semantics.} Engineering practice is driven by SLOs and contracts (e.g., ``90\% of utterances must be at least fair and show low disagreement''), not only average MOS.
\end{enumerate}

This paper proposes \textbf{contract-driven QoE auditing} that lifts ``scalar MOS regression'' to \emph{graph-level contract satisfaction}.
Given a deployment graph $G$ and a contract family $\mathcal{C}$, we report a vector $Q(G,\mathcal{C})$ of satisfaction rates that is interpretable, auditable, and more stable under common graph view transformations.

\subsection{Contributions (Hypotheses P0--P2)}
We summarize contributions as testable hypotheses:
\begin{itemize}
\item \textbf{P0: MOS regression as a special case.} We formalize experience contracts as Boolean predicates over rating vectors and show MOS regression corresponds to a degenerate contract set. Contract vectors reveal heterogeneity invisible to MOS.
\item \textbf{P1: View stability under graph transformations.} On SingMOS, contract-based quality exhibits smaller drift across system- vs.\ type-level views than MOS and graph-only baselines.
\item \textbf{P2: Contract semantics govern sample complexity.} On URGENT, learning difficulty is driven by semantic alignment between contracts and evidence, not just contract dimensionality.
\end{itemize}

\section{Related Work}
\subsection{MOS collection and standardization}
Subjective quality testing has long-standing standards.
ITU-T P.808 defines crowdsourcing procedures and controls for speech MOS \cite{ITU_P808}, and ITU-T P.800 specifies foundational subjective methods \cite{ITU_P800}.
For OTT streaming, ITU-T P.1203 provides parametric QoE estimation from bitstream and network traces \cite{ITU_P1203}.
These standards primarily target MOS estimation rather than explicit experience contracts; broader QoE perspectives are surveyed in \cite{QoEMoller}.

\subsection{Non-intrusive MOS prediction}
Deep non-intrusive MOS predictors achieve strong performance by regressing MOS from audio or representations.
NISQA predicts MOS and perceptual dimensions using neural architectures \cite{NISQA}.
VoiceMOS challenges benchmark self-supervised front-ends and transformer backbones \cite{VoiceMOS2022,VoiceMOS2024}.
Encoders such as wav2vec~2.0, HuBERT, and WavLM are widely used \cite{wav2vec2,HuBERT,WavLM}.
Most works optimize scalar MOS; explicit contract-level outputs and graph reasoning are rare.
Related efforts also study singing-voice quality assessment and bias correction in MOS-style settings \cite{Shi2024}, and learning-based QoE modeling for streaming scenarios \cite{BampisBovik}.
Challenge submissions and system reports further illustrate competitive modeling choices on VoiceMOS 2024, e.g., \cite{BabaT05}.

\subsection{Contracts, fairness, and graph views}
Contract-based design and SLO monitoring are common in distributed systems and service architectures \cite{ContractQoS,BorgOmegaK8s}.
Fairness constraints in ML are typically expressed as group-conditioned rate constraints \cite{EqualityOpp,FAwareness}.
Our work connects these perspectives by defining \emph{experience contracts} grounded in subjective rating vectors and evaluating them over \emph{service graphs}.

\section{Contract-Driven QoE Auditing}
\subsection{Service graph and rating model}
We represent a deployment as a labeled directed multigraph $G=(V,E)$.
Nodes may denote datasets, model variants, vocoders, post-filters, or user cohorts; an edge $e\in E$ corresponds to an evaluation episode.
Each edge carries a raw rating vector
$\mathbf{r}_e = (r_{e,1},\ldots,r_{e,J_e})$ with $r_{e,j}\in\{1,\ldots,5\}$,
derived statistics $\phi(\mathbf{r}_e)$ (mean, standard deviation, range, number of judges),
and optional non-intrusive features $\mathbf{x}_e$ (e.g., WavLM embeddings, metadata).

The conventional edge-level MOS is
\begin{equation}
\MOS(e)=\frac{1}{J_e}\sum_{j=1}^{J_e} r_{e,j}.
\end{equation}

\subsection{Experience contracts and contract vectors}
An \emph{experience contract} $c\in\mathcal{C}$ is a Boolean predicate over $\mathbf{r}_e$ and/or $\phi(\mathbf{r}_e)$:
\begin{align}
\varphi_{\text{lenient}}(e) &= \mathbb{I}\{\MOS(e)\ge 3.0\},\\
\varphi_{\text{strict}}(e)  &= \mathbb{I}\{\MOS(e)\ge 4.0\},\\
\varphi_{\text{fair}}(e)    &= \mathbb{I}\{\mathrm{std}(\mathbf{r}_e)\le 0.7 \wedge \mathrm{range}(\mathbf{r}_e)\le 2\},\\
\varphi_{\text{consensus}}(e) &= \varphi_{\text{lenient}}(e)\wedge \varphi_{\text{fair}}(e).
\end{align}

For a subset of edges $S\subseteq E$, the contract satisfaction vector is
\begin{equation}
Q(S,\mathcal{C})=\left(\frac{1}{|S|}\sum_{e\in S}\varphi_{c_k}(e)\right)_{k=1}^{K}\in[0,1]^K.
\label{eq:Q}
\end{equation}
We also use a scalar summary $Q_{\text{total}}$ as the mean of selected contract dimensions when needed.

\subsection{Graph views and view stability}
A \emph{graph view} is defined by a surjective mapping $\pi:V\rightarrow V'$ that induces a homomorphism $\pi: G\rightarrow G'$ by regrouping edges according to attributes (e.g., system ID, system type).
For view $v$, the induced edge partition is $\{S_i^{(v)}\}$ and each group has $Q_i^{(v)}=Q(S_i^{(v)},\mathcal{C})$.

To quantify view instability, we compute the mean absolute drift between two views (e.g., system vs.\ type):
\begin{equation}
\Delta_{\text{view}}(Q_k)=\E\left[ \left| Q_{k}^{\text{sys}} - Q_{k}^{\text{type}}\right|\right],
\end{equation}
and analogously for MOS. We report bootstrap confidence intervals when applicable.

\subsection{MOS regression as a special case (P0)}
Classical MOS regression learns $f_{\theta}:\mathbf{x}_e\mapsto \widehat{m}_e$ and evaluates via MAE/RMSE or correlation.
In our framework, it corresponds to a degenerate contract family $\mathcal{C}_{\MOS}=\{c\}$ with
$\varphi_c(e)=\mathbb{I}\{\MOS(e)\ge \tau\}$ for some threshold $\tau$, and the service graph restricted to utterance-level episodes.
Contract auditing retains multiple semantically grounded constraints and can therefore expose heterogeneity within the same MOS bin.

\section{Experimental Setup}
\subsection{Datasets}
Table~\ref{tab:datasets} summarizes the two public datasets used in this work.
URGENT2024\_MOS follows the URGENT dataset/challenge release \cite{URGENT}, and SingMOS v1 follows the dataset definition and metadata described in \cite{SingMOS}.

\begin{table}[t]
\caption{Datasets used in this work. $J$ is the average number of judges per utterance. ``Raw'' indicates per-utterance rating vectors are available.}
\label{tab:datasets}
\centering
\small
\setlength{\tabcolsep}{3pt}
\renewcommand{\arraystretch}{1.1}
\begin{tabularx}{\textwidth}{l l r r l >{\raggedright\arraybackslash}X}
\toprule
Dataset & Domain & \#Utts & \#Sys & Ratings & Inputs used in this work \\
\midrule
URGENT2024\_MOS & Speech QoE (mixed distortions, TTS/VC) & 6,900 & $\ge$40 & 1--5, $J{\approx}8$, raw &
WavLM utterance embeddings (mean+std pooling); waveforms not used for contract feature construction \\
SingMOS v1 & Singing QoE (ZH/JA) & 7,981 & 80 & 1--5, $J{\approx}5$, raw &
JSON rating/meta files only; audio not decoded \\
\bottomrule
\end{tabularx}
\end{table}

\subsection{Feature extraction}
For URGENT, we extract WavLM utterance embeddings using mean-pooled last-layer hidden states and use a mean+std pooling variant (concatenating first and second moments) as the main feature \cite{WavLM}.
We truncate utterances to 8 seconds for consistency.
Rating-derived statistics (mean, std, min, max, range, count) are used to instantiate contract labels, not as inputs to the MOS regressor.

For SingMOS, we load rating vectors from JSON metadata and compute utterance-, system-, and type-level aggregations.

\subsection{Models and baselines (URGENT)}
All URGENT models operate on WavLM mean+std embeddings.
\begin{itemize}
\item \textbf{MOS-only MLP:} two-layer ReLU MLP predicting scalar MOS with MSE loss.
\item \textbf{kNN memory:} embedding normalization + $k$NN (here $k{=}5$) in Euclidean distance; predicts MOS by averaging neighbors' MOS.
\item \textbf{C1 contract-aware auditor:} shared backbone with two heads: MOS regression and multi-label contract prediction over a chosen contract set $\mathcal{C}$. The contract loss is binary cross-entropy.
\item \textbf{Ablations:} (i) \emph{ID-contract} predicts a single contract-ID embedding rather than structured contracts; (ii) \emph{No-local-evidence} removes auxiliary evidence channels used to instantiate disagreement-related contracts, isolating the benefit of semantically aligned contract supervision.
\end{itemize}

We use three contract families: $\mathcal{C}_{\text{simple}}$ (lenient/strict), $\mathcal{C}_{\text{mid}}$ (lenient/strict/fair/consensus), and $\mathcal{C}_{\text{full}}$ (finer grid of thresholds).

\subsection{Training and metrics}
On URGENT, we use a fixed 20\% test set per seed and sample training fractions $\{0.1,0.2,0.5\}$ from the remainder (three seeds).
We report MOS MAE/RMSE.
For contracts we report AUPRC, F1, Brier score, and ECE; we also evaluate graph-level error
$\E\left[|Q_{\text{total}}^{\text{hat}}-Q_{\text{total}}^{\text{true}}|\right]$.

On SingMOS, we compute point estimates and 95\% bootstrap intervals (2,000 resamples) for MOS and contract drift under system- and type-level views.
Unless otherwise stated, SingMOS contract metrics use $\mathcal{C}_{\text{mid}}$ (lenient/strict/fair/consensus).

\paragraph{Graph-only baseline on SingMOS.}
In addition to raw MOS aggregation, we compute a graph-only baseline $\widehat{\MOS}$ (denoted as MOS\_hat in Fig.~\ref{fig:p1}) using only the released subjective ratings and metadata (no waveform decoding).
This baseline is not intended as a non-intrusive predictor; it serves to remove explicit contract semantics while retaining access to rating-derived summary statistics.
Concretely, we fit a regression model on system/type metadata and rating-derived summaries computed from the rating vectors, obtain $\widehat{\MOS}$ for each utterance, and then aggregate $\widehat{\MOS}$ under the same system- and type-level views as MOS before computing system--type drift with the same bootstrap procedure.

\section{Results}
\subsection{P0: MOS regression as a special case and explanatory gain}
Table~\ref{tab:urgent_main} shows that C1 improves MOS accuracy over the MOS-only MLP (0.318269 vs.\ 0.326217 MAE) while providing calibrated contract predictions.
The kNN memory baseline attains the lowest MOS MAE on this split (0.312301) but does not produce contract-level outputs and remains non-parametric, limiting its use for contract auditing and for modifying/adding new contracts.
Removing local evidence substantially degrades contract classification quality (e.g., F1@0.5 drops from 0.764221 to 0.590108) while leaving MOS MAE essentially unchanged relative to the MOS-only MLP, indicating that contracts encode additional semantics beyond scalar MOS.

\begin{table}[t]
\caption{URGENT (contract set $\mathcal{C}_{\text{mid}}$, train\_frac=0.5; averaged over seeds 13/21/42).}
\label{tab:urgent_main}
\centering
\footnotesize
\setlength{\tabcolsep}{3pt}
\renewcommand{\arraystretch}{1.15}
\begin{tabular}{@{}lccc cccc@{}}
\toprule
Model & Structured? & Local evid.? & MOS MAE$\downarrow$ & AUPRC$\uparrow$ & F1@0.5$\uparrow$ & Brier$\downarrow$ & ECE$\downarrow$ \\
\midrule
MOS-only MLP & No & -- & 0.326217 & -- & -- & -- & -- \\
C1 (full) & Yes & Yes & 0.318269 & 0.823765 & 0.764221 & 0.092962 & 0.035610 \\
ID-Contract ablation & ID only & Yes & 0.324725 & 0.820876 & 0.756428 & 0.093478 & 0.032222 \\
No-LocalEvidence ablation & Yes & No & 0.326217 & 0.790748 & 0.590108 & 0.102796 & 0.046231 \\
kNN memory ($k{=}5$) & No & -- & 0.312301 & -- & -- & -- & -- \\
\bottomrule
\end{tabular}
\end{table}

\begin{figure}[t]
\centering
\includegraphics[width=\linewidth]{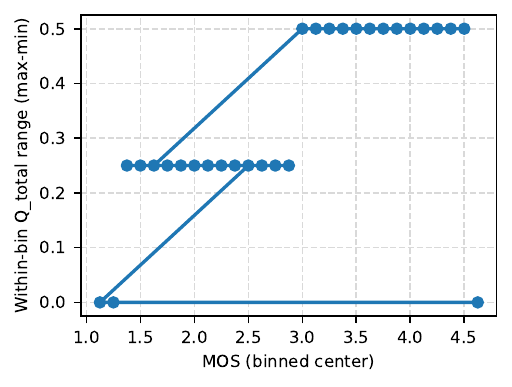}
\caption{Explanatory gain on URGENT: within MOS bins, contract satisfaction $Q_{\text{total}}$ can vary widely, revealing heterogeneity invisible to scalar MOS.}
\label{fig:p0}
\end{figure}

To quantify \emph{explanatory gain}, we bin utterances by MOS (step 0.125) and compute the within-bin range of $Q_{\text{total}}$.
Fig.~\ref{fig:p0} shows discrete and substantial within-bin heterogeneity in $Q_{\text{total}}$.
For MOS bin centers 1.375--2.875, the within-bin range is 0.25; for 3.000--4.500 it reaches 0.50, meaning utterances with the same MOS bin can differ by up to 50 percentage points in contract satisfaction.
At the extremes (e.g., bin centers 1.125, 1.250, and 4.625), the within-bin range collapses to 0.00.

\subsection{P1: View stability under graph transformations (SingMOS)}
\begin{figure}[t]
\centering
\includegraphics[width=\linewidth]{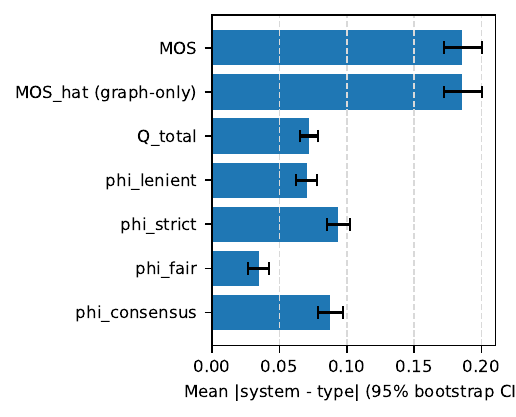}
\caption{SingMOS view stability: mean absolute system--type drift for MOS, a graph-only MOS baseline, and contract metrics (with 95\% bootstrap intervals). Contract-driven metrics drift less.}
\label{fig:p1}
\end{figure}

On SingMOS, we quantify view instability by the mean absolute system--type drift with 95\% bootstrap confidence intervals (2{,}000 resamples).
As shown in Fig.~\ref{fig:p1}, MOS exhibits a mean system--type drift of 0.185577 with a 95\% CI of [0.171933, 0.200236].
A graph-only baseline (MOS\_hat) remains essentially equally unstable: 0.185614 [0.172063, 0.200268].
In contrast, contract-driven metrics drift substantially less: $Q_{\text{total}}$ is 0.071624 [0.065104, 0.078937], and the disagreement-focused fairness contract is the most stable at 0.034768 [0.027013, 0.042562].
Other contracts lie between these extremes (e.g., $\varphi_{\text{strict}}$ 0.093486 [0.085122, 0.102435] and $\varphi_{\text{consensus}}$ 0.087653 [0.078844, 0.097358]).

\subsection{P2: Contract semantics govern sample complexity (URGENT)}
\begin{figure}[t]
\centering
\includegraphics[width=\linewidth]{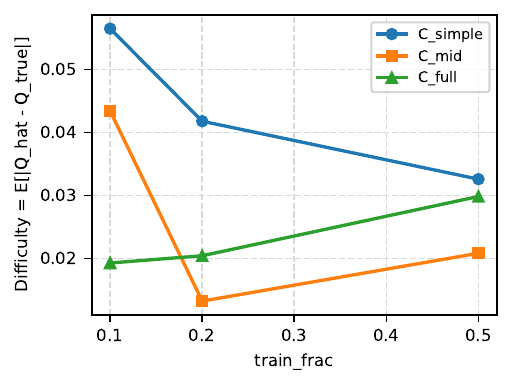}
\caption{Contract learning difficulty on URGENT vs.\ training fraction. Difficulty is governed by semantic alignment, not only $|\mathcal{C}|$.}
\label{fig:p2}
\end{figure}

To study sample complexity, we plot the difficulty
$D=\E\left[\left|Q_{\text{total}}^{\text{hat}}-Q_{\text{total}}^{\text{true}}\right|\right]$
for our main auditor (C1\_full) across training fractions and contract sets (Fig.~\ref{fig:p2}).
The curves are not necessarily monotonic in data and are not determined by $|\mathcal{C}|$ alone.
At train\_frac=0.1, $\mathcal{C}_{\text{simple}}$ is hardest ($D=0.056517$), while $\mathcal{C}_{\text{full}}$ is easiest ($D=0.019266$).
At train\_frac=0.2, $\mathcal{C}_{\text{mid}}$ achieves the lowest difficulty ($D=0.013211$), whereas $\mathcal{C}_{\text{full}}$ slightly increases ($D=0.020419$).
At train\_frac=0.5, $\mathcal{C}_{\text{simple}}$ remains relatively high ($D=0.032591$) and $\mathcal{C}_{\text{full}}$ increases further ($D=0.029838$), while $\mathcal{C}_{\text{mid}}$ rises to $D=0.020817$.
These results support that semantic alignment between contracts and evidence dominates effective sample complexity.

\section{Discussion and Limitations}
\textbf{Dataset coverage.} We evaluate one speech and one singing dataset; broader validation across telephony, conversational settings, and low-resource languages remains future work. \\
\textbf{Contract design space.} Current contracts are hand-crafted using MOS and simple disagreement statistics; extending to group-based fairness contracts and robustness constraints is a natural step \cite{EqualityOpp,FAwareness}. \\
\textbf{Graph structure.} We focus on view-induced partitions (utterance/system/type); extending to deeper multi-hop service graphs and GNN propagation is promising. \\
\textbf{Theory.} Formal generalization bounds for contract satisfaction under graph homomorphisms remain open.

\section{Conclusion}
We introduced a contract-driven QoE auditing framework that generalizes MOS regression to graph-level, contract-aware evaluation.
Across URGENT2024\_MOS and SingMOS, we show (P0) MOS regression is a special case that discards explanatory structure, (P1) contract metrics are more stable under view changes, and (P2) contract semantics dominate sample complexity.
The framework supports deployment-aligned, interpretable auditing and provides a foundation for contract-aware QoE evaluation in speech and multimedia services.

\end{document}